\RequirePackage{fix-cm}

\documentclass[twocolumn,epjc3]{svjour3}      
\smartqed  
\RequirePackage{amsmath}

\RequirePackage{amssymb}
\RequirePackage{mathrsfs}
\RequirePackage{amssymb,color}
\RequirePackage{amssymb,graphicx,amsmath}
\RequirePackage{float}
\RequirePackage{booktabs}

\RequirePackage{threeparttable}
\journalname{Eur. Phys. J. C}
\begin{document}

\title{The accelerated scaling attractor solution  of the  interacting agegraphic dark energy in
Brans-Dicke theory}

\author{Xian-Ming Liu \thanksref{e1,addr1,addr2}\and Zhong-Xu Zhai\thanksref{addr2} \and Kui Xiao \thanksref{addr3}\and Wen-Biao Liu (corresponding author)\thanksref{e2,addr2}}
\thankstext{e1}{e-mail: xianmingliu@mail.bnu.edu.cn}
\thankstext{e2}{e-mail: wbliu@bnu.edu.cn}

\institute{
Department of Physics, Hubei University for
Nationalities, Enshi Hubei, 445000, China \label{addr1}
\and
 Department of Physics, Institute of Theoretical Physics, Beijing Normal University, Beijing, 100875,
China\label{addr2}
\and
Department of Basic Teaching, Hunan Institute of Technology, Hengyang, 421002, China\label{addr3}
}

\date{Received: date / Accepted: date}

\maketitle

\begin{abstract}
We investigate the interacting agegraphic dark energy in Brans-Dicke
theory and introduce a new series general forms of dark sector
coupling. As examples, we select three cases involving a linear
interaction form (Model I) and two nonlinear interaction form (Model
II and Model III). Our conclusions show that the
accelerated scaling attractor solutions do exist in these models. We
also find that these interacting agegraphic dark energy modes are
consistent with the observational data. The difference in these
models is that nonlinear interaction forms give more approached
evolution to the standard $\Lambda$CDM model than the linear one.
Our work implies that the nonlinear interaction forms should be payed more attention.
\end{abstract}

\section{Introduction}

\label{sec:introduction}

Recent astronomical observations have provided strong evidence that our
universe is undergoing an accelerated expansion due to an exotic energy
component with negative pressure which is called dark energy \cite%
{Perlmutter-99,Bennett-03,Tegmark-04,Allen-04}. By far a leading candidate
of dark energy model is the Einstein's cosmological constant model ($\Lambda$%
CDM) which is consistent with the late-time observational data.
However, it is confronted with the so-called ``cosmological constant
problem" and ``coincidence problem" \cite{Weinberg-05}. Therefore,
many dynamical dark energy models \cite{Copeland-06}, such as
quintessence \cite{quintessence}, phantom \cite{phantom}, quintom
\cite{quintom}, tachyon \cite{tachyon}, generalized Chaplygin gas
\cite{Chaplygin}, etc, have also been taken into account.

The cosmological constant problem may be essentially an issue of
quantum gravity problem, since the cosmological constant is commonly
considered as the vacuum expectation value of some quantum fields.
Although a complete theory of quantum gravity has not been
established, according to some principles of quantum gravity, one
can make some attempts to probe the nature of dark energy model. The
holographic dark energy model is just an appropriate and interesting
example. This model is based on the holographic principle of quantum
gravity theory \cite{Hooft-93}, and is derived from the relationship
between the ultraviolet (UV) and the infrared (IR) cutoffs proposed
by Cohen {\emph{et al.}} in ref. \cite{Nelson-99}. According to the
limit set of the formation of a black hole, the UV-IR relationship
gives an upper bound on the zero point energy density
$\rho_q=3L^{-2}/8\pi G$, which means that the maximum entropy of the
system is of the order of $S^{3/4}_{BH} $, where $L$ is the scale of
IR cutoff and $S_{BH}$ is the entropy of the black hole.  Compared
with the holographic dark energy model and based on the
K$\acute{a}$rolyh$\acute{a}$zy relation  \cite{time-energy
uncertainty}, the so-called agegraphic dark energy model was
proposed, where the age of the universe $T=\int dt$ is used as the
IR cutoff $L$ \cite{Cai-07}. Furthermore, in ref.\cite{Cai-09}, the
interacting agegraphic dark energy has been introduced  and
investigated. It was shown that the equation of state of interacting
agegraphic dark energy can cross the phantom  division. The
interacting agegraphic dark energy model also has been extended to
the universe with spatial curvature in ref. \cite{Sheykhi-09}.
Recently, using the phase space analysis, it was shown that the
accelerated scaling attractor solutions of the interacting agegraphic
dark energy in the Einstein universe did exist and the results agree
with the observations \cite{Lemets-10}.

On the other side, scalar-tensor theories of gravity have been widely
applied to cosmology \cite{V.Faraoni-04}. The simplest alternative to
Einstein's general relativity which includes a scalar field in addition to
the tensor field is Brans-Dicke theory. This theory is more consistent with
the Mach's principle and less reliant on absolute properties of space \cite%
{Brans-Dicke-65}. It got a new impetus in recent years because it arises
naturally as the low energy limit of many theories of quantum gravity such
as super string theory or Kaluza-Klein theory. Noticing that the holographic
dark energy density belongs to a dynamical cosmology constant, it is more
natural for a dynamical frame to replace the Einstein's general gravity.
Therefore, it is worthwhile to investigate the holographic dark energy model
within the framework of Brans-Dicke theory \cite%
{Y.G.Gong-04,Y.G.Gong-08,Ahmad Sheykhi-09,J.Liu-10}. The extended
holographic dark energy model with Hubble horizon in Brans-Dicke theory has
been proposed and it is found that the model is not a viable dark energy
model unless the Brans-Dicke scalar field has a potential. So it is a very
interesting attempt to deeply investigate the agegraphic dark energy model in the
framework of Brans-Dicke theory .

Considering that dark energy (DE) and dark matter (DM) contribute to
the most fraction of the content of the universe, it is natural to
look into the possibility of the interaction between DE and DM,
which has been widely discussed \cite{D-D interaction}. It has been
argued that the coupling between DE and DM can provide a mechanism
to alleviate the coincidence problem and lead to an accelerated
scaling attractor solution with similar energy densities in the dark
sector today \cite{D-D interaction,Pavon-05,Boehmer-08,Chen Wang
Jing-08}. Noticing that there is no fundamental theory which can be
used to select a specific interacting dark energy model, any
interacting dark energy model will necessarily be phenomenological.
There are two criterions to determine whether the model is correct
and feasible. One is the observations, the other is to examine
whether the interacting model can lead to the accelerated scaling
attractor solutions, which is a decisive way to achieve similar
energy densities in dark sector and alleviate the coincidence
problem. In this work, we firstly introduce a new series of
interacting agegraphic dark energy models including linear and
nonlinear forms. Using the phase-plane analysis, it is found that
the accelerated scaling attractor solutions do exist in these
models. What's more, these agegraphic dark energy models are in
accordance with the late-time observational data.

Our  paper  is  organized  as  follows. In section \ref{Sec2}, the
agegraphic dark energy models in Brans-Dicke theory is constructed
and a series of dark sector coupling forms are introduced. In
section  \ref{Sec3}, using the phase-plane analysis, the accelerated
scaling attractor solutions are discussed in these models. In
section  \ref{Sec4}, using
the newly released Hubble parameter data \cite%
{Hubble..expand..Jimenez,Hubble..expand..Simon,Hubble..expand..Stern,Hubble..parameter..MaCong}%
, these agegraphic dark energy models are tested. Some conclusions
will be presented in section  \ref{Sec5}.

\section{Construction of the interacting agegraphic dark energy in
Brans-Dicke theory}

\label{Sec2} According to our metric convention, $(+,-,-,-)$, the Lagrangian
for Brans-Dicke theory with a scalar field in the Jordan frame is
\begin{eqnarray}
\mathcal{L}_{BD}=\sqrt{-g}[(-\varphi R+\omega \frac{1}{\varphi}%
g^{\mu\nu}\partial_\mu\varphi\partial_\nu\varphi)+\mathcal{L}_M(\Psi)],
\label{1.0}
\end{eqnarray}
where the dimensionless $\omega$ is the coupling constant, and $\mathcal{L}%
_M(\Psi)$ is the matter Lagrangian. The current observational constraint on $%
\omega$ is $\omega>10^4$; it recovers Einstein's general relativity when $%
\omega\rightarrow\infty$. In particular, it is expected that $\varphi(t,%
\overrightarrow{x})$ is spatially uniform and evolves slowly only with
cosmic time $t$ so that $\varphi(t,\overrightarrow{x})\rightarrow \varphi(t)$%
. Following ref. \cite{Arik-08}, we can introduce a new field $\phi$
as
\begin{equation}
\varphi=\frac{1}{8\omega}\phi^2 .  \label{1.1}
\end{equation}
So the Lagrangian for Brans-Dicke theory can be written as following form
\begin{equation}
\mathcal{L}_{BD}=\sqrt{-g}[-\frac{1}{8\omega}\phi^2 R+\frac{1}{2}%
g^{\mu\nu}\partial_\mu\phi\partial_\nu\phi+\mathcal{L}_M(\Psi)] .
\label{1.2}
\end{equation}

Considering a classical perfect fluid with the energy-momentum tensor $%
T^\mu_\nu=diag(\rho,-p,-p,-p)$, the gravitational field equations
derived from the variation of the action Eq. (\ref{1.2}) with
respect to the flat Robertson-Walker metric are
\begin{eqnarray}
&\ & \frac{3}{4\omega}\phi^2H^2-\frac{1}{2}\dot{\phi}^2+\frac{3}{2\omega}H%
\dot{\phi}\phi=\rho ,  \label{1.3} \\
&\ & -\frac{1}{4\omega}\phi^2(2\dot{H}+3H^2)-\frac{1}{\omega}H\dot{\phi}%
\phi-(\frac{1}{2}+\frac{1}{2\omega})\dot{\phi}^2-\frac{1}{2\omega}\ddot{\phi}
\phi=p,  \label{1.3.1} \\
&\ &\ddot{\phi}+3H\dot{\phi}=\frac{3}{2\omega}(\dot{H}+2H^2)\phi,
\label{1.3.2} \\
&\ &\dot{\rho}+3H(\rho+p)=0,  \label{1.3.3}
\end{eqnarray}
where the dot is the derivative with respect to time and $H=\frac{\dot{a}}{a}
$ is the Hubble parameter. Combining the above equations, we get
\begin{equation}
-\dot{H}=-4\left(\frac{H\dot{\phi}}{\phi}\right)+2\omega\left(\frac{\dot{\phi%
}}{\phi}\right)^2+\frac{4\omega}{(2\omega+3)\phi^2}[\omega p+(\omega+2)\rho].
\label{1.3.4}
\end{equation}

The total matter is supposed to be composed of two parts: pressureless dark
matter and agegraphic dark energy. So the total energy density $\rho$
includes the energy density of agegraphic dark energy $\rho_q$ and the
energy density of dark matter $\rho_m$ in this model. In Brans-Dicke theory,
the effective gravitational constant $G_{\mathrm{eff}}$ can be defined as $%
G_{\mathrm{eff}}^{-1}=\frac{2\pi}{\omega}\phi^2$, the agegraphic dark energy
can naturally be defined as
\begin{equation}
\rho_q=\frac{3n^2\phi^2}{4\omega T^2},
\end{equation}
where $n$ is a positive constant. Now we turn to considering the interaction
between dark matter and dark energy. The balance equations of the agegraphic
dark energy and dark matter can be written respectively as \cite{Ahmad
Sheykhi-09}
\begin{eqnarray}
&\ &\dot{\rho}_q+3H(1+\omega_q)\rho_q=-Q,  \label{inte1} \\
&\ &\dot{\rho}_m+3H\rho_m=Q,  \label{inte2}
\end{eqnarray}
where $Q$ denotes the phenomenological interaction term. The interacting
term is always a function of the Hubble parameter $H$, the density of dark
energy, and the density of the dark matter. One can often find the following
interaction model forms \cite{Lemets-10,Cai-09,Chimento-10,D-D
interaction,Pavon-01,Cai-05,Zhang-05,Zhangxin2011,Ahmad Sheykhi-09}
\begin{eqnarray}
Q=&\ &3H\alpha\rho_q,\, 3H\beta\rho_m,\,3\gamma
H(\rho_q+\rho_m),\notag\\
&\ &3H(\alpha\rho_q+\beta\rho_m),\,3\beta H
\rho_q^\alpha\rho_m^{1-\alpha} ,  \label{l-1}
\end{eqnarray}
where $\alpha,\beta,\gamma$ are positive constant parameters. We would like
to consider the more general interaction as
\begin{equation}
Q=3H\rho_q g(\xi)=3H\rho_q \sum^{+\infty}_{i=-\infty}A_i\xi^i,
\end{equation}
where $\xi=\frac{\rho_m}{\rho_q}$ , $g(\xi)=
\sum^{+\infty}_{i=-\infty}A_i\xi^i$, and $A_i$ is a positive constant
parameter with respect to $\xi^i$. Then Eq.(\ref{inte1}) and Eq.(\ref{inte2}%
) can be written as
\begin{eqnarray}
&\ &\dot{\rho}_q+3H(1+\omega_q)\rho_q=-3H\rho_q g(\xi),  \label{c-1} \\
&\ &\dot{\rho}_m+3H\rho_m=3H\rho_q g(\xi) .  \label{c-2}
\end{eqnarray}
In this paper, as examples, $g(\xi)$ will be selected as:
\begin{eqnarray}
&\text{Model}\, \text{I:} &g_1(\xi)=\alpha_1+\beta_1 \xi ,  \label{m-1} \\
&\text{Model}\, \text{II:} &g_2(\xi)=\alpha_2+\beta_2 \xi^2 ,  \label{m-2} \\
&\text{Model}\, \text{III:} &g_3(\xi)=\alpha_3\xi^{-1}+\beta_3 \xi .
\label{m-3}
\end{eqnarray}
It is obvious that Model I is linear interaction, while Model II and III are
nonlinear interactions, which can be taken as the extending of Eq. (\ref{l-1}%
).

\section{Phase-space analysis}

\label{Sec3}

According to the discussion in section  \ref{Sec2}, it is easy to see that eqs. (%
\ref{1.3}), (\ref{1.3.2}), (\ref{1.3.4}), (\ref{c-1}), and (\ref{c-2}) can
give a closed system which can determine the cosmic behavior. In order to
study the dynamical behavior of interacting dark energy and dark matter, we further introduce the following dimensionless variables
\begin{equation}
x=\sqrt{\frac{4\omega}{3\phi^2H^2}\rho_q},\, y=\sqrt{\frac{4\omega}{%
3\phi^2H^2}\rho_m},\, \lambda=\frac{\dot{\phi}}{H\phi}.
\end{equation}
Similar forms of selecting these dimensionless variables in
agegraphic dark energy model was firstly used in ref.
\cite{Lemets-10}. Using these dimensionless variables, Eq. (\ref%
{1.3}) becomes
\begin{equation}
x^2+y^2+\frac{2\omega}{3}\lambda^2-2\lambda=1 .  \label{2.1}
\end{equation}
Now, the critical density can be defined as $\rho_c=\frac{3H^2}{8\pi G_{%
\mathrm{eff}}}=\frac{3\phi^2H^2}{4\omega} $ \cite{J.Liu-10}, so Eq. (\ref{2.1}%
) becomes $\Omega_q+ \Omega_m+ \Omega_\phi=1$, where
\begin{equation}
\Omega_q=x^2,\, \Omega_m=y^2,\, \Omega_\phi=\frac{2\omega}{3}%
\lambda^2-2\lambda .  \label{o-1}
\end{equation}
From Eq. (\ref{o-1}), we find that the Brans-Dicke scalar field $\phi$
plays a role of dark energy, so we can assume that both the
agegraphic and the scalar field drive our universe to accelerate
\cite{J.Liu-10}. So we have
\begin{equation}
\Omega_{DE}=\Omega_q+\Omega_\phi,
\end{equation}
then
\begin{equation}
r=\frac{\Omega_{DM}}{\Omega_{DE}}=\frac{\Omega_m}{\Omega_q+\Omega_\phi}=%
\frac{1}{x^2+\frac{2\omega}{3}\lambda^2-2\lambda}-1.\label{rate}
\end{equation}
Noticing that $0\leq\Omega_q, \Omega_\phi, \Omega_m\leq1$, the limit of $x$,
$y$, and $\lambda$ can be obtained as $0\leq x\leq 1$, $0\leq y\leq 1$, $%
\frac{3}{2\omega}(1-\sqrt{1+\frac{2\omega}{3}})\leq\lambda\leq0$ or $\frac{3%
}{\omega}\leq\lambda\leq\frac{3}{2\omega}(1-\sqrt{1+\frac{2\omega}{3}}) $
respectively.

Subsequently, thinking of Eqs. (\ref{1.3})- (\ref{1.3.4}), we have
\begin{eqnarray}
    &\ &x'=-x\left[f(x,y,\lambda)+\frac{x}{n}\right],\label{2-1}\\
     &\ &y'=-y\left[\frac{3}{2}(1-\frac{x^2g(\xi)}{y^2})+\lambda+f(x,y,\lambda)\right]\label{2-2},\\
      &\ &\lambda'=-3\lambda-\lambda^2+\frac{3}{\omega}+\left(\frac{3}{2\omega}-\lambda\right)f(x,y,\lambda),\label{2-3}
\end{eqnarray}
where $x^{\prime }=dx/dN,y^{\prime }=dy/dN,\lambda ^{\prime }=d\lambda
/dN,N=\ln a,\xi =\frac{y^{2}}{x^{2}}$, and
\begin{eqnarray}
f(x,y,\lambda )&=&\frac{\dot{H}}{H^{2}}=4\lambda -2\omega \lambda ^{2}-\frac{%
3\omega }{2\omega +3}[1-g(\xi )-\frac{2}{3}\lambda +\frac{2}{3c}x]x^{2}\notag\\
&-&
\frac{3(\omega +2)}{2\omega +3}(1-\frac{2\omega }{3}\lambda ^{2}+2\lambda ).
\label{2-4}
\end{eqnarray}%
The state parameter $\omega _{q}$ for the agegraphic dark energy could be
expressed in terms of these new variables as
\begin{equation}
\omega _{q}=-1+\frac{2}{3n}x-\frac{2}{3}\lambda -g(\xi ).
\end{equation}%
For completeness, we give the deceleration parameter
\begin{equation}
q=-\frac{\ddot{a}}{aH^{2}}=-1-\frac{\dot{H}}{H^{2}}=-1-f(x,y,\lambda ).
\end{equation}%
In the interacting agegraphic dark energy models, the properties of
agegraphic dark energy are determined by the parameters $A_{i}$, $\omega $
and $n$. Eqs. (\ref{2-1})-(\ref{2-4}) are just the functions of $x,y,$ and $%
\lambda $, not the functions of $N$ or other variables. So this dynamical
system is just a three-dimensional autonomous system. For an autonomous
system $\mathbf{X}^{\prime }=f(\mathbf{X})$, there are some critical points $%
\mathbf{X}_{c}$ satisfying $\mathbf{X}^{\prime }=0$, so we have
\begin{eqnarray}
&\ &-x_{c}(f(x_{c},y_{c},\lambda _{c})+\frac{x_{c}}{n})=0,  \label{3-1} \\
&\ &-y_{c}[\frac{3}{2}(1-\frac{x_{c}^{2}g(\xi _{c})}{y_{c}^{2}})+\lambda
_{c}+f(x_{c},y_{c},\lambda _{c})]=0,  \label{3-2} \\
&\ &-3\lambda _{c}-\lambda _{c}^{2}+\frac{3}{\omega }+(\frac{3}{2\omega }%
-\lambda _{c})f(x_{c},y_{c},\lambda _{c})=0,  \label{3-3}
\end{eqnarray}%
where $\xi _{c}=\frac{y_{c}^{2}}{x_{c}^{2}}$. In order to determine the
stability property of the critical points, it is necessary to expand the
autonomous system $\mathbf{X}^{\prime }=f(\mathbf{X})$ around the critical
points $\mathbf{X}_{c}$. Setting $\mathbf{X}=\mathbf{X}_{c}+\mathbf{U}$, where $\mathbf{U}$ is a column vector of the perturbation of the variables,
one can expand the equation for the perturbation up to the first order as $%
\mathbf{U}^{\prime }=\mathbf{M}\cdot \mathbf{U}$, where the matrix $\mathbf{M%
}$ contains the coefficients of the perturbation equations. The stability
property for each critical point is determined by the eigenvalues of $\mathbf{M}
$ \cite{Xiao-IJMPA}. The $3\times 3$ matrix $\mathbf{M}$ of the linearized
perturbation equations is
\begin{eqnarray*}
\mathbf{M}&=&
\left(
\begin{array}{c}
 -x(\frac{\partial f}{\partial x}+\frac{2}{n})\\
  -y(-\frac{3x}{y^{2}}g(\xi )+\frac{3}{x}\frac{\partial g(\xi )}{\partial \xi }%
+\frac{\partial f}{\partial x}) \\
(\frac{3}{2\omega }-\lambda )\frac{\partial f}{\partial x}
\end{array}
\right.\\&\ &
\begin{array}{c}
  -x\frac{\partial f}{\partial y} \\
  -[\frac{3}{2}+\lambda +f+\frac{3x^{2}}{%
2y^{2}}g(\xi )-3\frac{\partial g(\xi )}{\partial \xi }+y\frac{\partial f}{%
\partial y}] \\
(\frac{3}{
2\omega }-\lambda )\frac{\partial f}{\partial y}
\end{array}
\\&\ &
\left.
\begin{array}{c}
 -x\frac{\partial f}{\partial \lambda } \\
-y(1+\frac{\partial f}{\partial \lambda }) \\
-3-2\lambda -f+(\frac{3}{%
2\omega }-\lambda )\frac{\partial f}{\partial \lambda }%
\end{array}
\right)
,
\end{eqnarray*}%
where
\begin{eqnarray*}
&\ &\frac{\partial f}{\partial x}=-\frac{6\omega }{2\omega +3}\left[ \frac{%
x^{2}}{n}-xg(\xi )+\frac{y^{2}}{x}\frac{\partial g(\xi )}{\partial \xi }%
\right] , \\
&\ &\frac{\partial f}{\partial y}=y\frac{6\omega }{2\omega +3}\frac{\partial
g(\xi )}{\partial \xi }, \\
&\ &\frac{\partial f}{\partial \lambda }=4(1-\omega \lambda )-\frac{6(\omega
+2)}{2\omega +3}\left( 1-\frac{2}{3}\omega \lambda \right) +\frac{2\omega }{%
2\omega +3}x^{2}, \\
&\ &\frac{\partial g(\xi )}{\partial \xi }=\sum_{i=-\infty }^{+\infty
}iA_{i}\xi ^{i-1}.
\end{eqnarray*}%
We can examine the sign of the real part of the eigenvalues of $\mathbf{M}%
_{x=x_{c},y=y_{c},\lambda =\lambda _{c}}$, which determines the type and
stability of the critical points $(x_{c},y_{c},\lambda _{c})$.
The simplest finite critical points and their properties for this model are summarized in the Table \ref{tab:attractor}.
\begin{table*}[htbp]
 \centering \small
 \begin{threeparttable}
 \caption{\label{tab:attractor} Location of the critical points of the autonomous system of Eqs.(\ref{2-1}), (\ref{2-2}) and (\ref{2-3}), their stability and dynamical behavior of the Universe at those points.}
  \begin{tabular}{lccc}
   \hline
 $(x_c,y_c,\lambda_c)$ coordinates & Stability  character &   q   \\
\hline
 $(0,0,\lambda_1=\frac{3}{2\omega}(1\pm\sqrt{1+\frac{2\omega}{3}}))$ & unstable &$-4-4\lambda_1$ \\

   $(0,y_2=\sqrt{1-\frac{2\omega}{3}\lambda_2^2+2\lambda_2},\lambda_2=\frac{3(\omega+1)}{8\omega}[1\pm\sqrt{1+\frac{56\omega}{3(\omega+1)^2}}])$& unstable &$-1-\frac{6\omega\lambda_2+\omega\lambda_2-6}{3-2\omega\lambda_2}$ \\
 $(x_3,0,\lambda_3=\frac{3}{2\omega}[1\pm\sqrt{1+\frac{2\omega}{3}(1-x_3^2)}])$& unstable & $-1+\frac{x_3}{n}$\\
    $(x_\ast=\frac{n(6\omega\lambda_2^2+\omega\lambda_2-6)}{3-2\omega\lambda_2},y_\ast=\sqrt{1-\frac{2\omega}{3}\lambda_\ast^2+2\lambda_\ast}-x_\ast^2,\lambda_\ast)$& attractor &$-1+\frac{x_\ast}{n}$\\
 \hline
  \end{tabular}
  \small
  \end{threeparttable}
\end{table*}
Here we only concentrate on the attractor, so the real part of the eigenvalues of $%
\mathbf{M}_{x=x_{\ast},y=y_{\ast},\lambda =\lambda _{\ast}}$ should be negative.
If  the autonomous system of Eqs.(\ref{2-1}), (\ref{2-2}) and (\ref{2-3}) presents scaling solutions, the coincidence problem gets substantially alleviated because, regardless of the initial conditions, the system evolves toward a final state where the ratio of dark matter to dark enenrgy
stays constant. Using Eq. (\ref{rate}), it is easy to find that
\begin{equation}
  r'=-(1+r^2)[2x x'+(\frac{4\omega}{3}\lambda-2)\lambda']
\end{equation}
The scaling solutions mean $r'=0$. Obviously the attractor $(x_\ast,y_\ast,\lambda_\ast)$ is just a scaling solution, because $x'|_{x_\ast}=0,\lambda'|_{\lambda_\ast}=0$ lead to $r'|_{x_\ast,y_\ast,\lambda_\ast}=0$.
What's more, if the critical point is the accelerated scaling attractor
solution, the following constraints $q(x_{\ast},y_{\ast},\lambda
_{\ast})<0,r(x_{\ast},y_{\ast},\lambda _{\ast})\approx \frac{0.26}{0.74}\approx 0.353$
should also be satisfied.

\begin{figure}[tbp]
\includegraphics[width=0.5\textwidth]{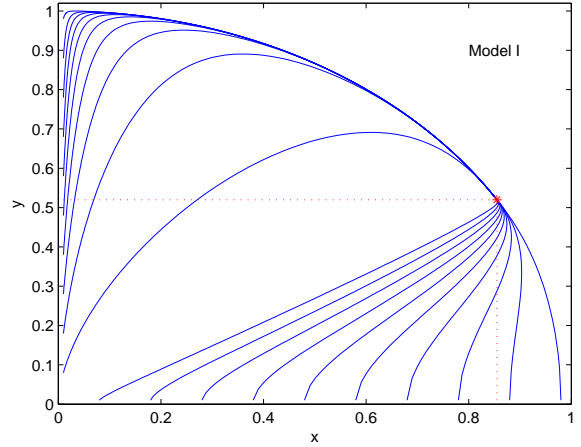}
\caption{The phase plane for the model I with $n=20$, $\protect\omega=10^{3}
$, $\protect\alpha_{1}=0.3$ and $\protect\beta_{1}=0.16$. The red star
stands for the late-time attractor with $x_{\ast}\approx0.86$, $y_{\ast}\approx0.26
$, $\protect\lambda_{\ast}\approx9.9\times10^{-4}$.}
\label{fig:model1}
\end{figure}

\begin{figure}[tbp]
\includegraphics[width=0.5\textwidth]{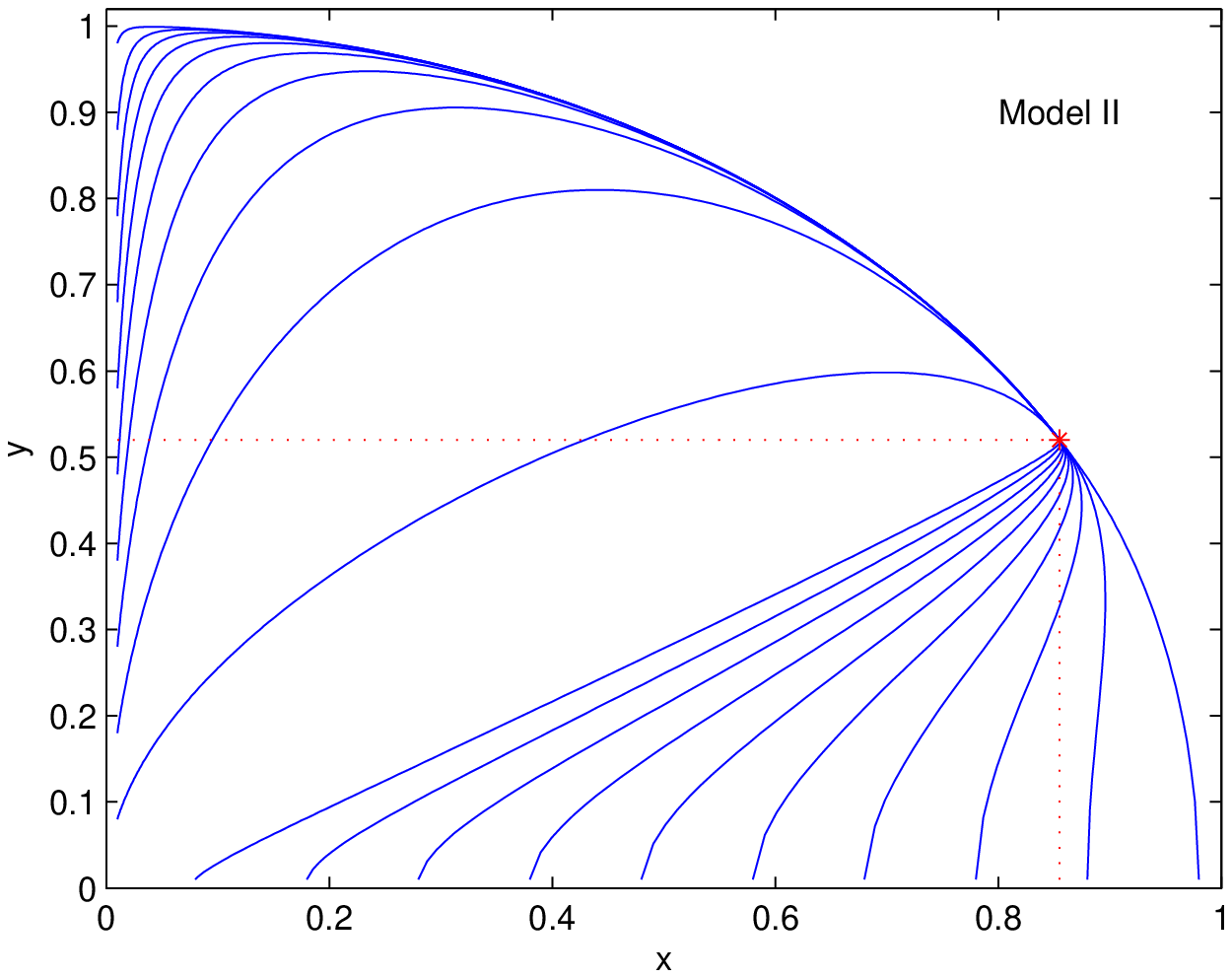}
\caption{The phase plane for the model II with $n=20$, $\protect\omega%
=10^{3}$, $\protect\alpha_{2}=0.36$ and $\protect\beta_{2}=10^{-6}$. The
red star stands for the late-time attractor with $x_{\ast}\approx0.86$, $%
y_{\ast}\approx0.26$, $\protect\lambda_{\ast}\approx9.9\times10^{-4}$.}
\label{fig:model2}
\end{figure}

\begin{figure}[tbp]
\includegraphics[width=0.5\textwidth]{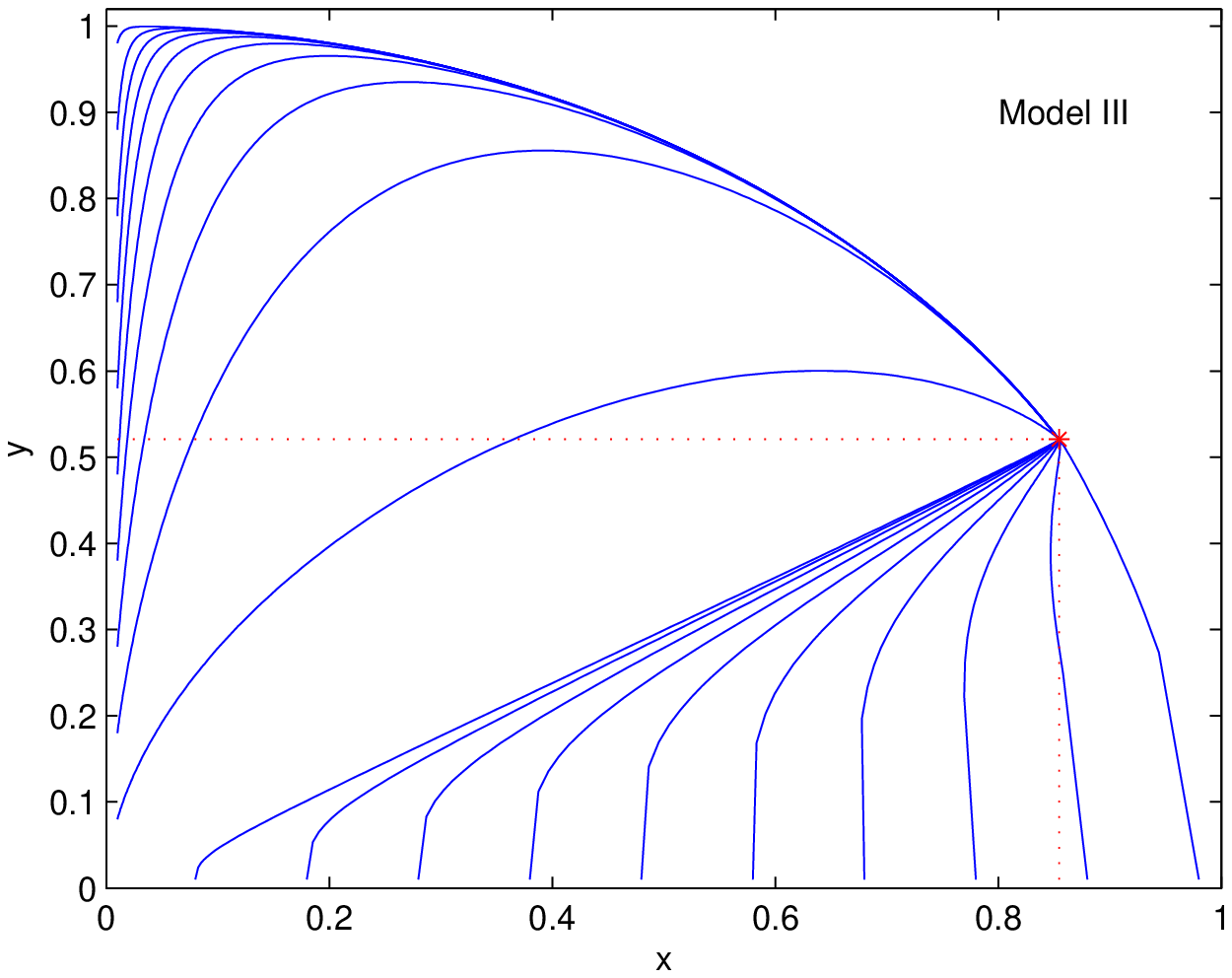}
\caption{The phase plane for the model III with $n=20$, $\protect\omega%
=10^{3}$, $\protect\alpha_{3}=0.12$ and $\protect\beta_{3}=0.1$. The red
star stands for the late-time attractor with $x_{\ast}\approx0.86$, $%
y_{\ast}\approx0.26$, $\protect\lambda_{\ast}\approx9.9\times10^{-4}$.}
\label{fig:model3}
\end{figure}

Obviously the accelerated scaling attractor solutions can be obtained by fixing
the parameters $A_i$, $\omega$, and $n$. We have found the accelerated
scaling attractor solutions in Model I, Model II, and Model III using the
numerical analysis method. The results are shown in Fig. \ref{fig:model1}, %
\ref{fig:model2}, \ref{fig:model3} respectively. The numerical solutions
show that the cosmic evolution is insensitive to the initial conditions. The
different interacting model forms just affect the intermediate evolution.
This makes the coincidence problem to be substantially alleviated.
\begin{figure}[tbp]
\includegraphics[width=0.5\textwidth]{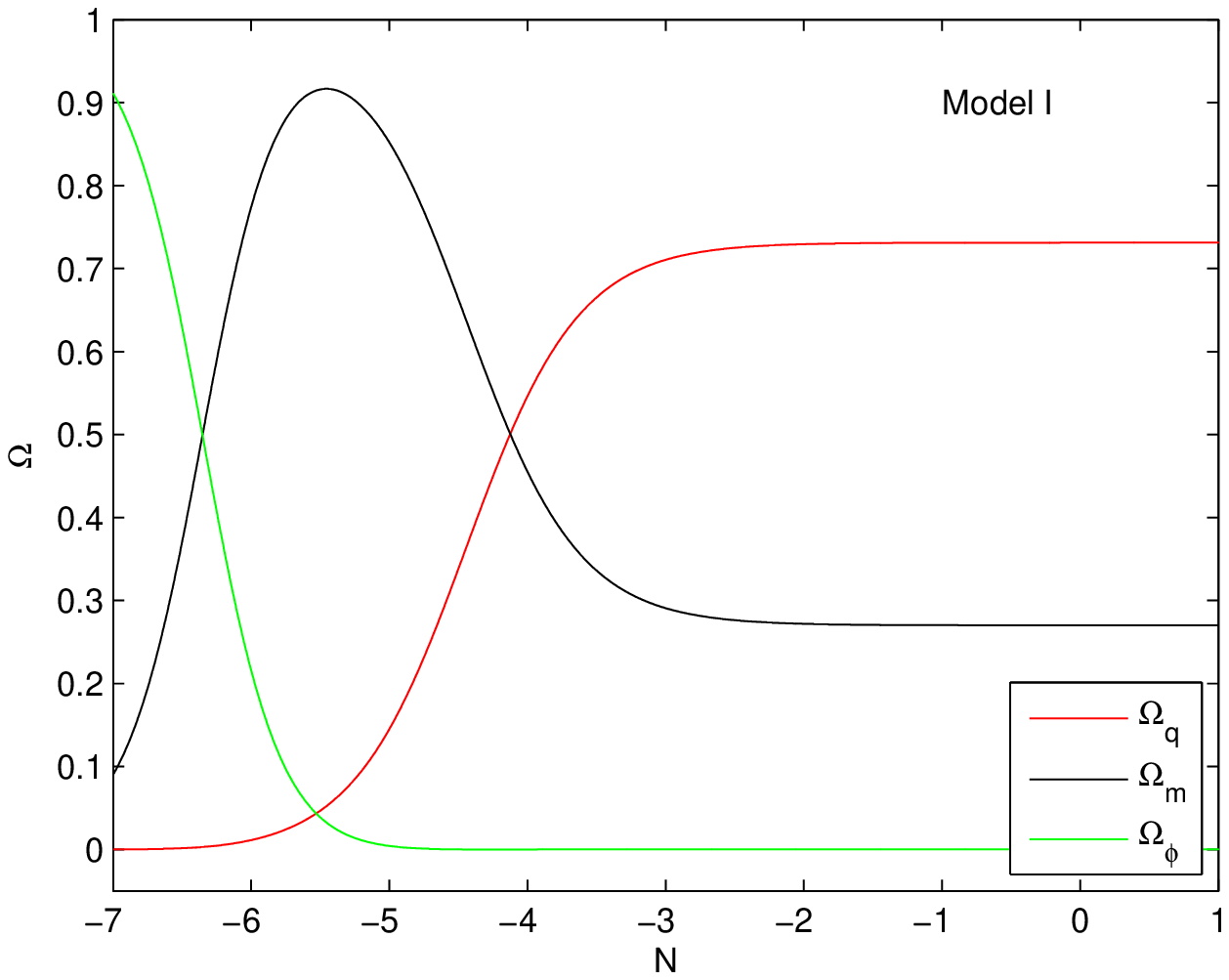} %
\includegraphics[width=0.5\textwidth]{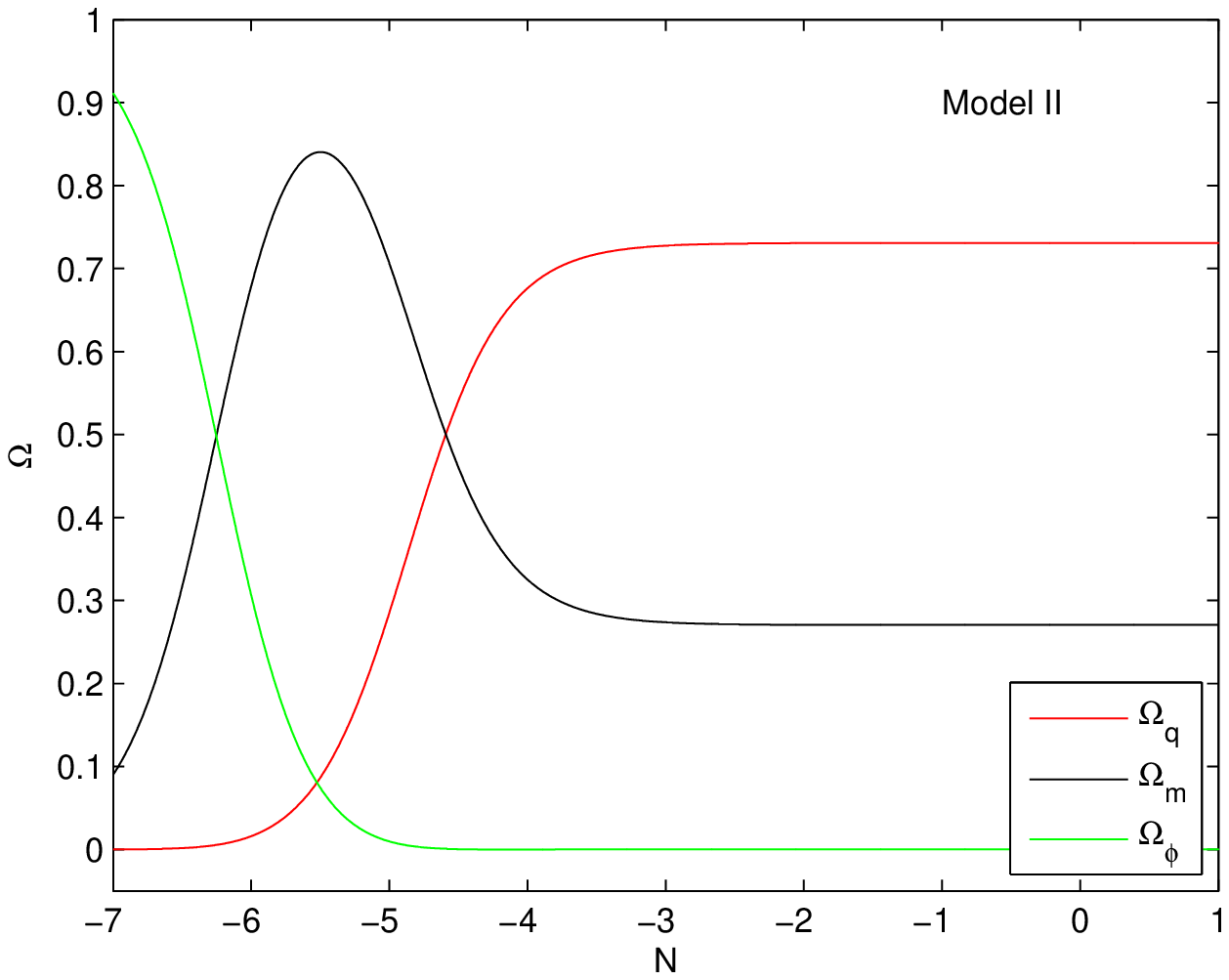} %
\includegraphics[width=0.5\textwidth]{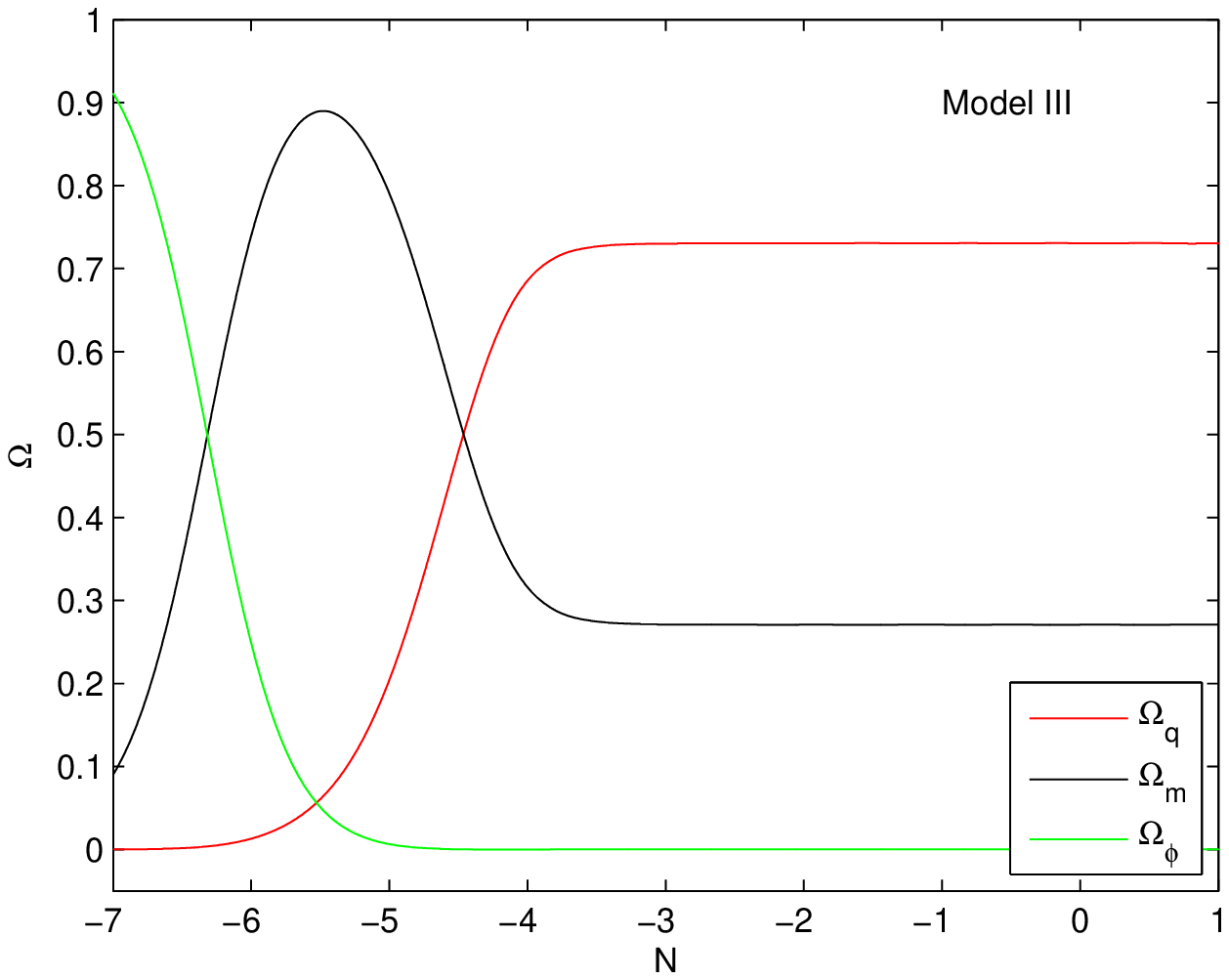}
\caption{Behavior of $\Omega_{m}$(black), $\Omega_{q}$(red) and $\Omega_{%
\protect\phi}$(green) as a function of $N=\ln a$ for the particular models}
\label{fig:density}
\end{figure}

\begin{figure}[tbp]
\includegraphics[width=0.5\textwidth]{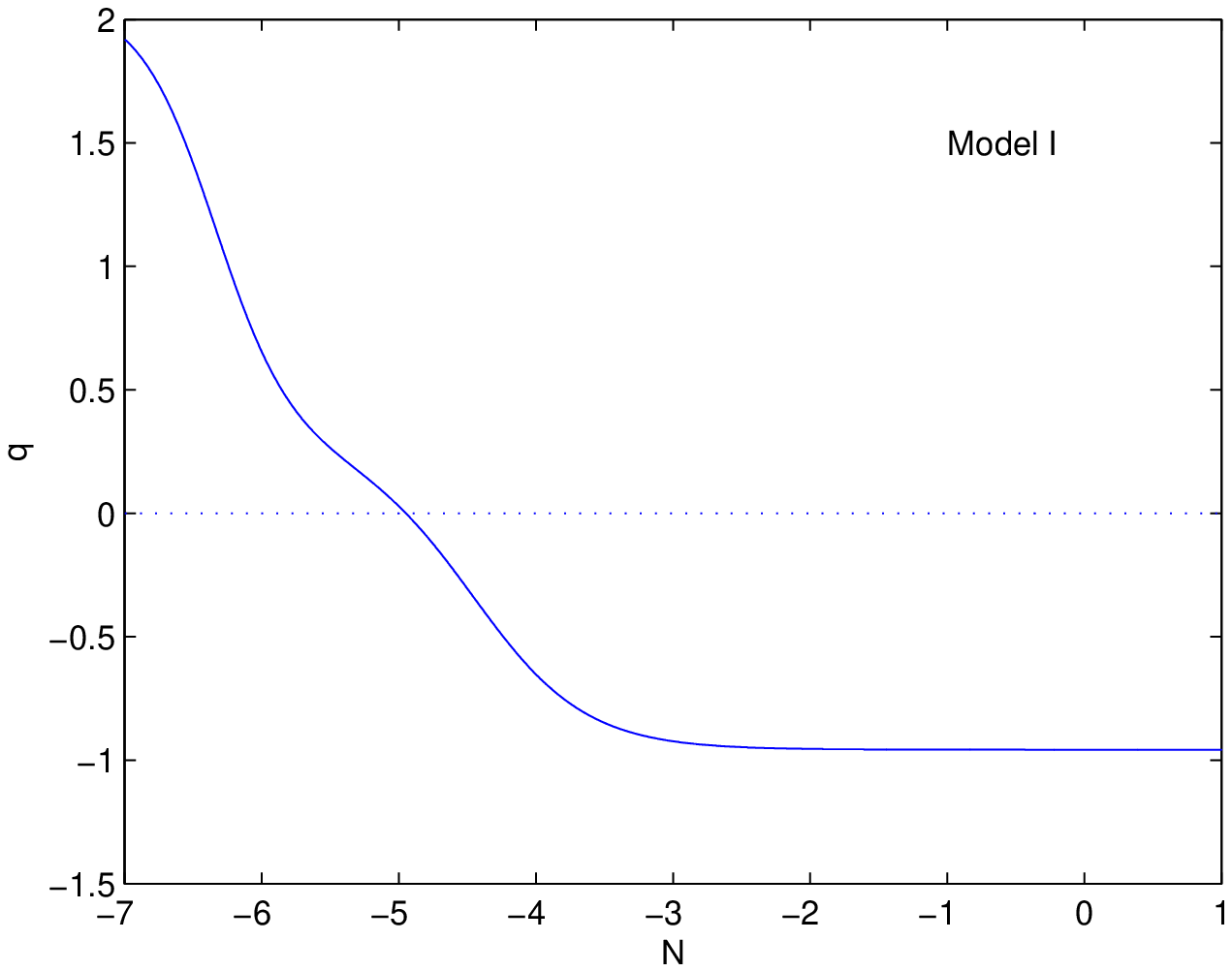} \includegraphics[width=0.5%
\textwidth]{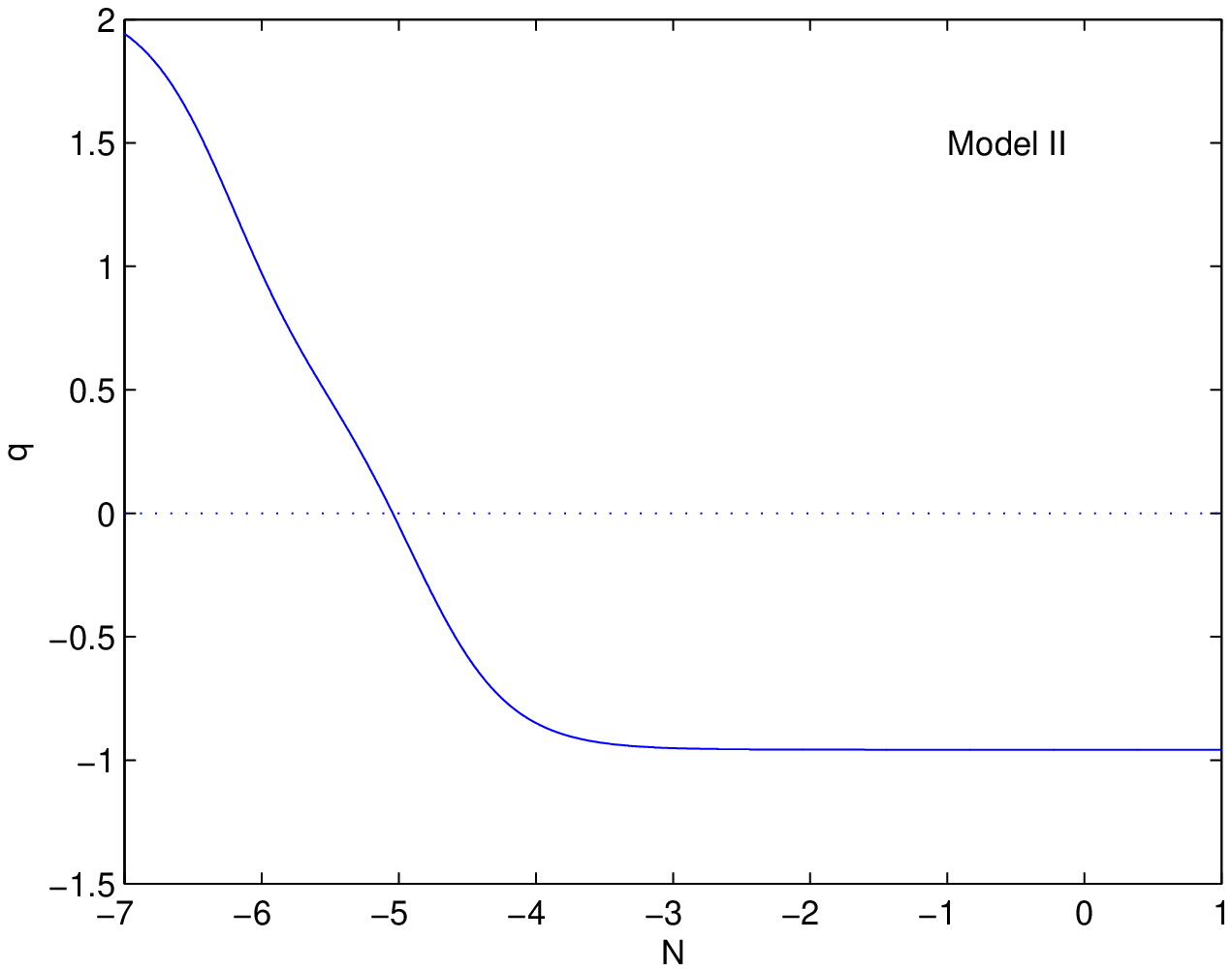} \includegraphics[width=0.5\textwidth]{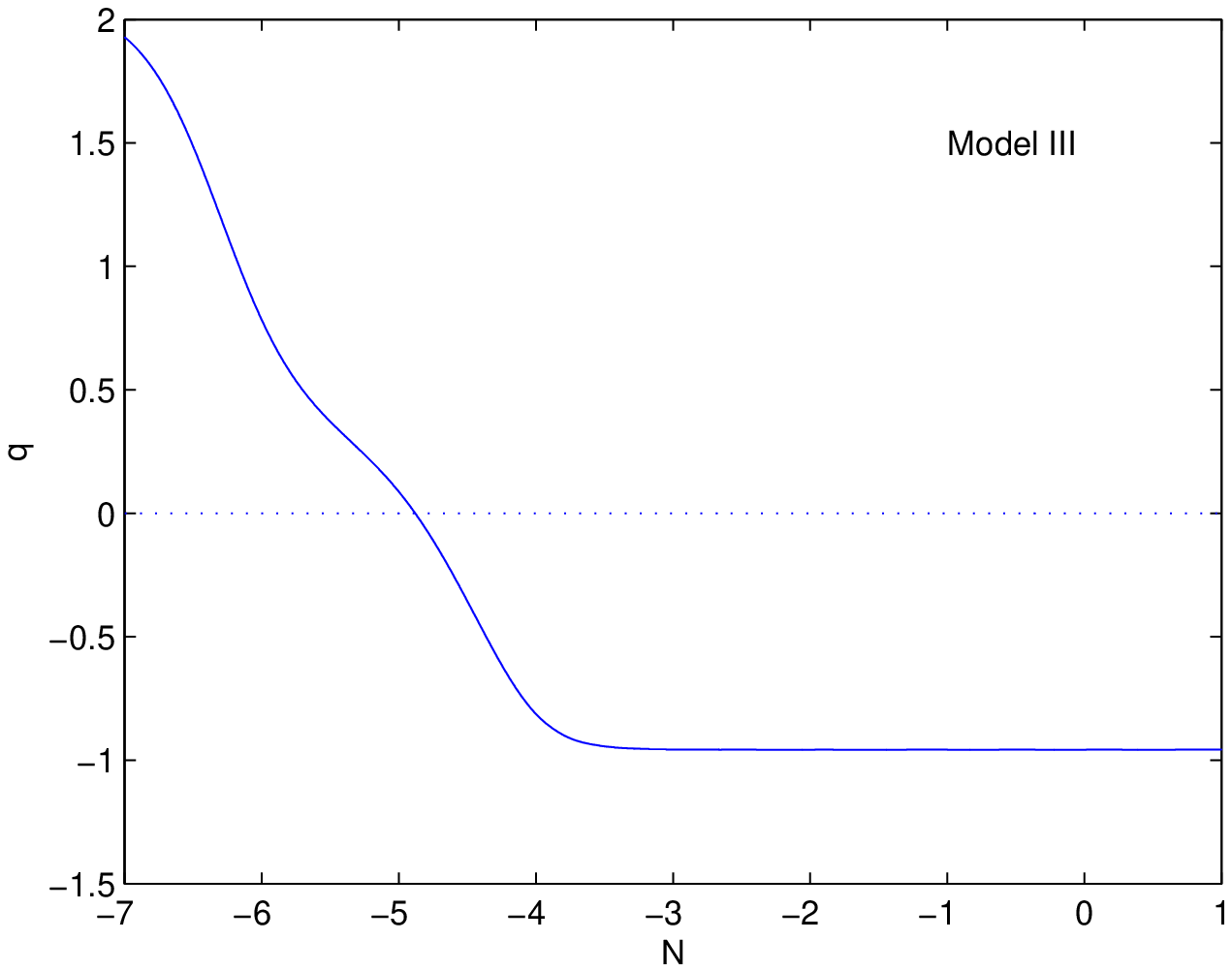}
\caption{Behavior of deceleration parameter $q$ as a function of $N=\ln a$
for the particular models}
\label{fig:q}
\end{figure}
\begin{figure}[tbp]
\includegraphics[width=0.5\textwidth]{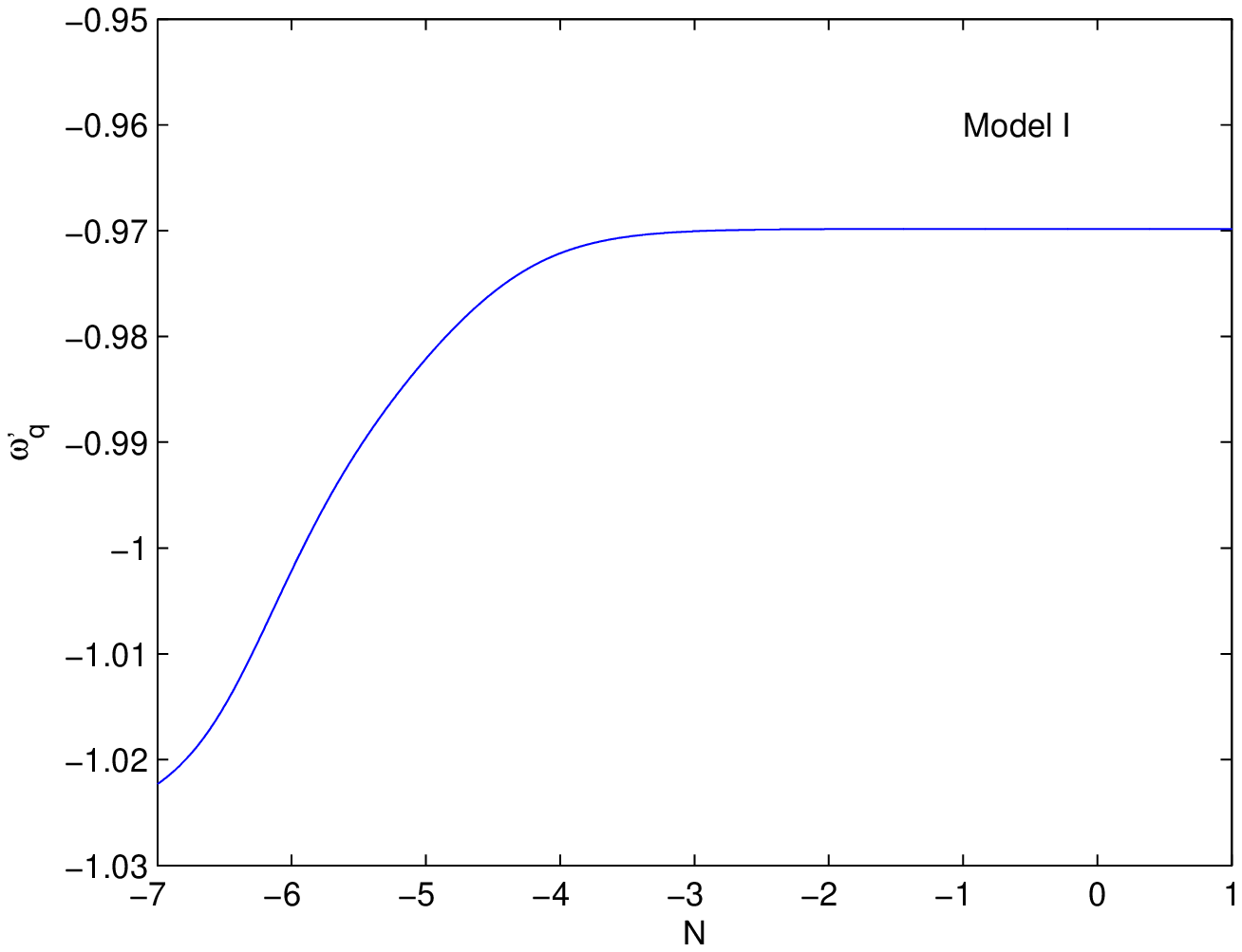} \includegraphics[width=0.5%
\textwidth]{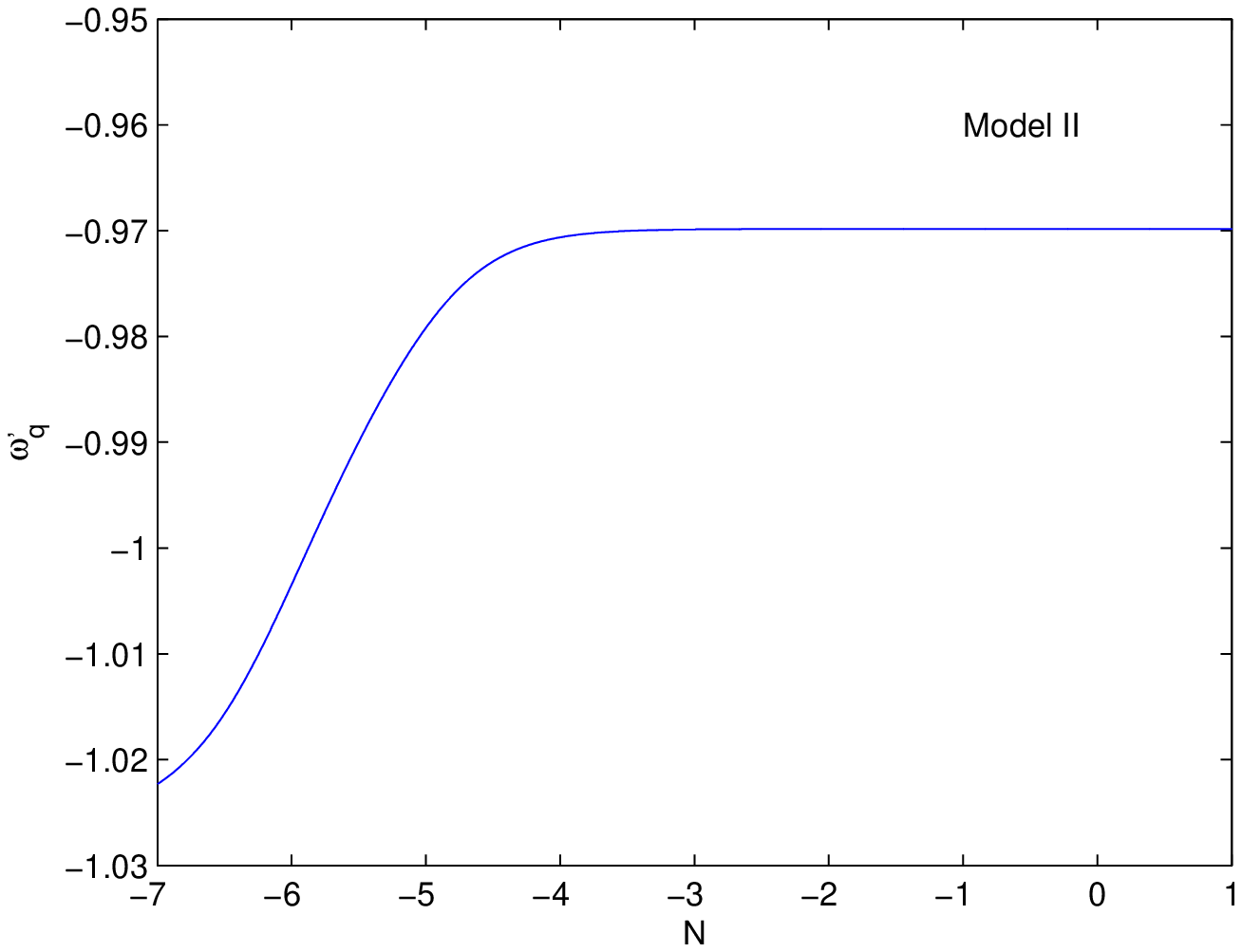} \includegraphics[width=0.5\textwidth]{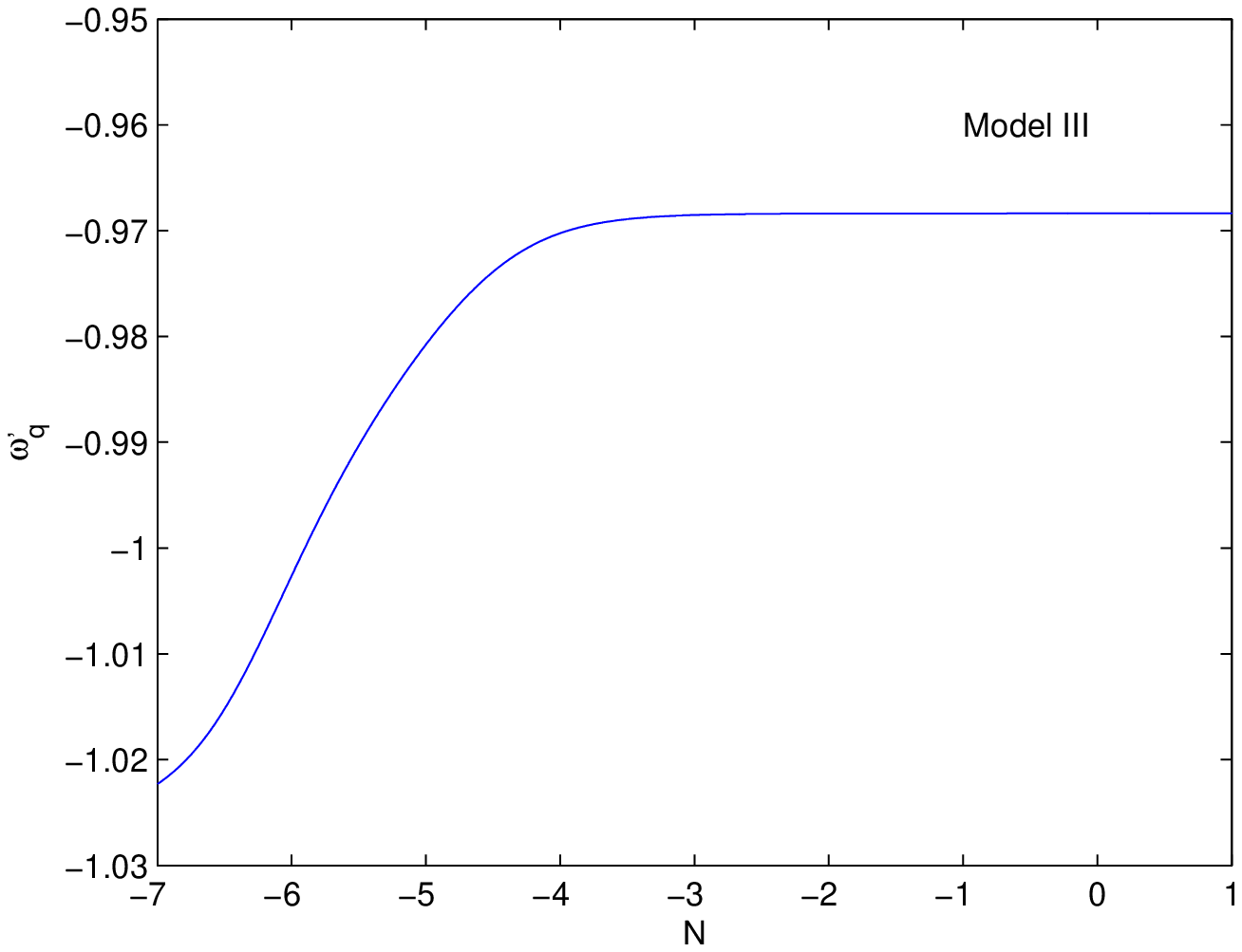}
\caption{Behavior of effective equation of state of dark energy $\protect%
\omega_q^{\prime }$ as a function of $N=\ln a$ for the particular models}
\label{fig:EOS}
\end{figure}
Figures \ref{fig:density}, \ref{fig:q}, \ref{fig:EOS} show that the
three interacting agegraphic dark energy models give very similar
conclusions. From Fig. \ref{fig:density}, one can see the cosmic
evolution in these models, where $\Omega_\phi$ dominates the
universe at the early epoch and then the matter is becoming
increasingly dominant over $\Omega_\phi$ afterwards. After a
transitory matter-dominated epoch, the universe becomes accelerated
expansion in the agegraphic dark energy dominated epoch. From Fig.
\ref{fig:q}, the universe is now undergoing an accelerated expansion
phase. The similar curves of $q$ in three models show that this
cosmic acceleration is arising recently. The evolution of equations
of state of dark energy is presented in Fig. \ref{fig:EOS}. There
is a phantom-like to quintessence-like transition while the values
of $\omega_{q}^{^{\prime }}$ is always around -1.

\section{Observational data}

\label{Sec4}

Now, we would like to check the three interacting agegraphic dark
energy models using the observational data and compare them with the
standard $\Lambda $CDM model.

The continuity eqs. (\ref{c-1}) and (\ref{c-2}) can also be written
in the standard forms
\begin{eqnarray}
&\ &\dot{\rho}_q+3H(1+\omega^{\prime }_q)\rho_q=0,  \label{4.1} \\
&\ &\dot{\rho}_m+3H(1+\omega^{\prime }_m)\rho_m=0,  \label{4.2}
\end{eqnarray}
where $\omega^{\prime }_{q}=-1-\frac{2}{3}(\lambda-\frac{x}{n}%
),\,\omega^{\prime }_{m}=\frac{g(\xi)}{\xi}$.

The Brans-Dicke field can be assumed as a  power law of the scale factor\cite%
{Ahmad Sheykhi-09}
\begin{equation}
\phi=\phi_0a^{\lambda}.  \label{4.3}
\end{equation}
Noticing that the scalar field $\Omega_\phi$ in figure
\ref{fig:density} is very infinitesimal
in the late-time universe, we can assume that $\lambda$ is a constant $%
\lambda\approx\lambda_c$, and the product $\lambda_c \omega$ results
order unity, which is consistent with refs.\cite{Ahmad
Sheykhi-09,N.Banerjee-07}.

Substituting Eqs. (\ref{4.1}), (\ref{4.2}), and (\ref{4.3}) into Eq. (\ref{1.3}%
), one can get the evolution of Hubble parameter $H=H_0E(z)$, where the
expansion rate $E(z)$ can be expressed as
\begin{eqnarray}
&\ &E^2(z)=(1+2\lambda_c-\frac{2\omega}{3\lambda_c^2})^{-1}(1+z)^{2\lambda_c}\notag\\
&&[\Omega_{m0} \exp(3\int^z_0\frac{1+\omega^{\prime }_{m}}{1+z^{\prime }}%
dz^{\prime }) +\Omega_{q0}\exp(3\int^z_0\frac{1+\omega^{\prime }_{q}}{%
1+z^{\prime }}dz^{\prime })]\notag\\,
\end{eqnarray}
where the subscript "0" denotes the present value.

\begin{figure}[tbp]
\includegraphics[width=0.45\textwidth]{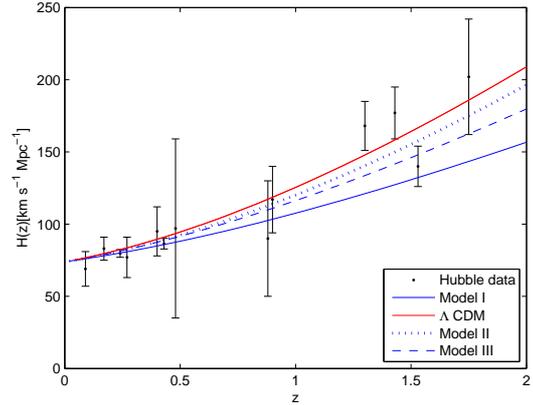}
\caption{The theoretical evolution of $H(z)$ with observational data points
(black): $\Lambda$CDM (red solid), Model I (blue solid), Model II (blue
dotted), and Model III (blue dashed)}
\label{fig:data}
\end{figure}

We will use the newly released Hubble parameter data \cite%
{Hubble..expand..Jimenez,Hubble..expand..Simon,Hubble..expand..Stern,Hubble..parameter..MaCong,Zhai}
which is directly related to the expansion history of the universe
by its definition: $H=\dot{a}/a$ to determine a preferable
interacting model. In the meantime, we choose the standard
$\Lambda$CDM model as the fiducial model which fits the observations
best. The results are presented in Fig. \ref{fig:data}.

Apparently, the nonlinear interaction forms (Model II and Model III) give
better evolutions to the standard $\Lambda$CDM one. The linear one shows
evident deviation from the $\Lambda$CDM and observation data. The nonlinear
interaction between dark energy and dark matter is better to describe the
real physical process. This comparison is incomplete and it is only a rough
overview. It is necessary to further explore the nonlinear interaction
between the dark section of the universe.

\section{Discussions and conclusions}

\label{Sec5}

Holographic dark energy model is an interesting attempt to
investigate the nature of dark energy in the framework of quantum
gravity. Considering that the simplest alternative to Einstein's
general relativity which includes a scalar field in addition to the
tensor field is Brans-Dicke theory and the holographic bound can be
satisfied for both the $k=0$ and $k=-1$ universe in the Brans-Dicke cosmology\cite{Gongprd60}, it is natural to extend the
research to the holographic dark energy models of this theory. In this paper, we have investigated
the  interacting agegraphic dark energy in the flat ( $k=0$ )
Brans-Dicke cosmology.
Firstly, a series of new general forms of dark sector coupling are
introduced and the accelerated scaling attractor solutions have been
found. Moreover, using the newly released Hubble parameter data, we
have also tested these interacting agegraphic dark energy models.

The interacting term can be selected as a function of $H$, the
density of dark energy, and the density of the dark matter.
According to this requirement, we have proposed the general
interacting agegraphic dark energy models. Three cases including a
linear interaction form (Model I) and two nonlinear interaction
forms (Model II and Model III) have been investigated. Using the
phase-plane analysis, the dynamical behavior of these models has
been investigated and it was found that the accelerated scaling
attractor solutions did exist in these models. This can alleviate the
coincidence problem.

Afterwards, using the newly released Hubble parameter data, we tested these
dynamical dark energy models. These interacting
agegraphic dark energy modes have given a series of reasonable
pictures of the cosmic evolution and they were consistent with the
late-time observational data. In particular, we found the nonlinear interaction forms (Model II and Model
III) gave more approached evolution to the standard $\Lambda$CDM
model than the linear one (Model I). Our work show that we should pay more attention to the nonlinear interaction forms rather than the linear form. This deserves further investigations.
\begin{acknowledgements}
The authors would like to thank the anonymous referee
for his/her insightful and constructive suggestions, which allowed us to improve the manuscript
significantly.
This work is supported by the National Natural Science
Foundation of China under Grant Nos. 10773002, 10875012, and 11175019. It is
also supported by the Fundamental Research Funds for the Central
Universities under Grant No. 105116.
\end{acknowledgements}



\end{document}